\documentclass[11pt,a4paper]{article}
\usepackage{graphicx} \usepackage{epsfig} 
\usepackage{subeqn} 

\newcommand{\C}{{\mathcal C}}

\newcommand{\ZB}{{Z_{\rm B}}}
\newcommand{\ZF}{{Z_{\rm F}}}

\newcommand{\dbk}{d_{{\rm B},k}}
\newcommand{\e}{{\rm e}}
\newcommand{\eb}{\epsilon_{\rm B}}

\newcommand{\ebi}{\epsilon_{{\rm B},i}}
\newcommand{\ebj}{\epsilon_{{\rm B},j}}
\newcommand{\ebk}{\epsilon_{{\rm B},k}}
\newcommand{\ebo}{\epsilon_{\rm B,0}}
\newcommand{\ebp}{\epsilon_{\rm B}^\prime}
\newcommand{\ebki}{\epsilon^{i}_{{\rm B},k}}
\newcommand{\ef}{\epsilon_{\rm F}}
\newcommand{\ei}{\epsilon_i}

\newcommand{\kb}{k_{\rm B}}
\newcommand{\mub}{\mu_{\rm B}}

\newcommand{\rmd}{{\rm d}}

\newcommand{\sib}{\sigma_{\rm B}}

\newcommand{\sibeps}{\sigma^{\epsilon_{{\rm B},i}}_{\rm B}}

\newcommand{\EB}{{E_{\rm B}}}
\newcommand{\EF}{{E_{\rm F}}}
\newcommand{\SB}{{S_{\rm B}}}
\newcommand{\SF}{{S_{\rm F}}}

\newcommand{\UB}{U_{\rm B}}
\newcommand{\ugs}{U_{\rm gs}}
\newcommand{\wbk}{w_{{\rm B},k}}
\newcommand{\xb}{x_{\rm B}}
\newcommand{\xbp}{x_{\rm B}^\prime}
\newcommand{\bX}{{\bf X}}

\begin{document}

\begin{center}
{\Large THE THERMODYNAMIC EQUIVALENCE AT WORK: TRANSFORMING FERMI SYSTEMS INTO BOSE SYSTEMS}

\bigskip \bigskip

\small
DRAGO\c S-VICTOR ANGHEL

\bigskip

\footnotesize
{\em Department of Theoretical Physics, NIPNE--HH\\ Str. Atomistilor no.407, P.O.BOX MG-6\\ Bucharest-Magurele, Romania\\ E-mail: dragos@theory.nipne.ro}

\bigskip

\small

(Received \today ) 
\end{center}

\bigskip

\footnotesize

{\em Abstract.\/} 
We generalize the method introduced in J. Phys. A: Math. Gen. {\bf 35}, 
7255 (2002) based on the concept of \textit{thermodynamic equivalence} 
and we transform a Fermi system of general density of states 
into a thermodynamically equivalent Bose system. This consists of mapping 
configurations of fermions from the original system onto configurations of 
bosons, the initial and final configurations having 
the same energy above the many-body ground state energy. In this way we 
obtain two systems of particles of different exclusion statistics, but 
which have the same entropies--and therefore identical 
canonical thermodynamic properties. 
This method enables one in general to calculate the system properties 
in either of the bosonic and fermionic representations. 
We check the method here in microscopic detail by calculating the 
equilibrium particle distributions 
in the two representations using the entropy maximization 
at fixed particle number and fixed ``fermionic'' and ``bosonic'' energies, 
respectively. 
Analytical calculations seem difficult to do, but we check the 
results numerically and we find them equal within the numerical accuracy. 

\bigskip


\normalsize

\section{INTRODUCTION \label{intro}}

The heat capacity of ideal systems of constant density of single 
particle states (DOS) and fixed number of particles is independent of 
the exclusion statistics of the constituent particles 
\cite{ProcCambrPhilos42.272.1946.Auluc,PhysRev.135.A1515.1964.May,AmJPhys63.369.Viefers,PhysLettA212.299.1996.Isakov,PhysRevLett.74.3912.1995.Sen,PhysRevE.55.1518.1997.Lee,PhysRevE.56.4854.1997.Apostol,JPA35.7255.2002.Anghel}. 
Moreover, integrating 
the ratio between the heat capacity, $\C$, and the temperature, $T$, 
with respect to the temperature, we obtain the entropy, 
$S=\int_0^T (\C/T')dT'$. So $S$, expressed as a function of $T$, particle 
number $N$, and all the other extensive parameters of the system (let us 
denote them by $\bX$), is also 
independent of the exclusion statistics. Based on this property, systems 
of equal, constant, DOS have been called 
\textit{thermodynamically equivalent} \cite{PhysRevE.55.1518.1997.Lee}. 

The thermodynamic equivalence relation can be extended from the set of 
systems of constant DOS to the set of all physical systems 
\cite{JPA35.7255.2002.Anghel}. To define the 
thermodynamic equivalence from a more general perspective, 
let me split the internal enegy of the system as $U\equiv \ugs+\UB$, 
where $\ugs$ is the energy of the system at zero temperature and $\UB$ 
is the \textit{excitation} or \textit{Bose energy}. Then two systems of 
entropies 
$S_1(U_1,N,\bX)$ and $S_2(U_2,N,\bX)$ are thermodynamically equivalent 
if $S_1(\UB+U_{{\rm g.s.},1}(N,\bX),N,\bX)=S_2(\UB+U_{{\rm g.s.},2}(N,\bX),N,\bX)$. 
The thermodynamic equivalence splits the set of all physical system into 
equivalence classes--like the classes which include the ideal systems of 
the same, constant DOS \cite{JPA35.7255.2002.Anghel,JPA36.L577.2003.Anghel}. 

The equality of the entropies of two systems with the same $\UB$ and 
$(N,\bX)$ implies that the number of microstates with the same excitation 
energy is the same in the two systems 
for any values of $\UB$ and extensive parameters $(N,\bX)$. 
This was shown explicitly to be true for systems of the same, constant DOS, 
by mapping configurations of particles of arbitrary exclusion statistics 
onto configurations of bosons of the same $\UB$ 
\cite{JPA35.7255.2002.Anghel}. 
Vice-versa, the mapping of configurations of ideal bosons 
into configurations of ideal fermions puts into correspondence the 
bosons on the ground state (the Bose-Einstein condensate) 
with a degenerate subsystem of fermions at the 
bottom of the single particle spectrum, which was called the 
{\em Fermi condensate} \cite{JPA35.7255.2002.Anghel,JPA36.L577.2003.Anghel}. 
The configurations of fermions which contain the Fermi condensate have higher 
probability than the ``standard'' Fermi distribution, as it was shown 
in Ref. \cite{JPA36.L577.2003.Anghel}, contrary to what was 
generally belived. In interacting systems, the existence of the Fermi 
condensate may modify so dramatically the weights of different 
configurations of particles that it may lead to first order phase 
transitions 
\cite{JPA35.7255.2002.Anghel,RomRepPhys59.235.2007.Anghel}.

In this paper we generalize the method of Ref. \cite{JPA35.7255.2002.Anghel} 
and we show how to map a Fermi system of 
general DOS to a thermodynamically equivalent Bose system. The 
technique may be extended (although not in a trivial way) to any 
transformation between Bose, Haldane, and 
Fermi systems and, to distinguish it from other transformations, it will be 
called {\em exclusion statistics transformation} (EST).

In Section \ref{BoseDOS} we calculate the quasiparticle energies and 
the density of states of the Bose gas. If the Fermi DOS is not constant, 
the Bose quasiparticle energies depend on the population of the other 
quasiparticle levels; in other words, the bosons are interacting. 
This effect is discussed in Section \ref{BoseInt}. 

By EST, the particle population of the energy levels in the Bose system 
is determined by the particle population in the Fermi system and vice-versa. 
On the other hand, once we set the rules for calculating the DOS and the 
interaction between the particles in the Bose gas, the particle populations 
in the two systems (Bose and Fermi) can be 
calculated independently, by entropy maximization. 
In this way it is shown that one can make calculations in either of the 
systems and transfer the results to the other one. 

Last section is reserved for the discussion of the results 
and conclusions.

\section{THE TRANSFORMATION FROM FERMI TO BOSE SYSTEMS
\label{iEST}}

\begin{figure}[t]
\begin{center}
\unitlength1mm\begin{picture}(50,45)(0,0)
\put(0,0){\psfig{file=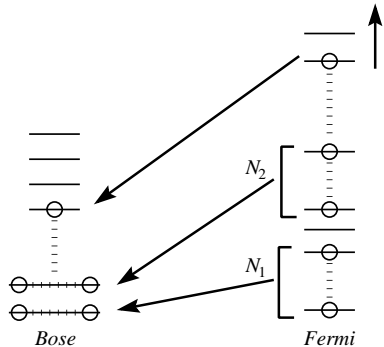,width=50mm}}
\end{picture}
\caption{The illustration of the basic idea of exclusion statistics transformation (EST).}
\label{total_B}
\end{center}
\end{figure}

Concretely, at microscopic scale the EST from a Fermi to a Bose system 
may be imagined as follows: we picture the fermions 
as forming groups of ``close packed'' particles 
that occupy completely certain intervals along the single particle energy axis 
(no empty single particle states in the intervals) and we associate 
to each of these groups an energy level in the 
equivalent Bose system (see figure \ref{total_B}). The groups are labeled 
with $i=0,1,\ldots$, starting from the bottom of the energy axis and going up. 
So, the group $i$ of $N_i$ fermions is associated to the level $\ebi$ 
of $N_i$ bosons, 
in such a way that $\epsilon_{{\rm B},0}=0$, $\ebi\le\ebj$ if $i<j$, and 
the total energy, $\EB$, of the new system is the same 
as the excitation energy of the original Fermi system--the 
\textit{excitation energy} of the Fermi system being the difference between 
the energy of the system, $\EF$, and the lowest energy attainable at the 
given particle number, $\ugs$ (in other words, $\ugs$ is the energy of the 
system at zero K). 

To distinguish between the energies of individual configurations and the 
(average) internal energy of the 
system in given thermodynamic conditions, we denote by 
$\EF$ the energy of a configuration of fermions (the Fermi 
energy will be denoted by $\ef$) and by $\EB(=\EF-\ugs)$ the energy of a 
configuration of bosons. Since for a given excitation energy, 
$\EB$, to each configuration of fermions it corresponds a configuration 
of bosons and vice-versa, then the entropy of the Fermi system, $\SF$, of 
internal energy $U=\ugs+\EB$, is equal to the entropy of the Bose 
system of internal energy 
$\UB=\EB$ [$\SF(\ugs+\UB,N,\ldots)=\SB(\UB,N,\ldots)$], i.e. the two systems 
are \textit{constructed to be thermodynamically equivalent}. 
With these definitions we have 
$T^{-1}=\partial S_{\rm B}/\partial E_{\rm B}=\partial S_{\rm F}/\partial E_{\rm F}$ and $\mub/T=-\partial S_{\rm B}/\partial N=(\mu-\ef)/T$, where 
$\mu/T=-\partial S_{\rm F}/\partial N$ and 
$\ef\equiv \partial U_{\rm g.s.}/\partial N$ \cite{JPA35.7255.2002.Anghel}. 
In figure \ref{total_B} we see also how the Fermi condensate 
\cite{JPA36.L577.2003.Anghel} becomes Bose-Einstein condensate in the 
Bose system \cite{JPA35.7255.2002.Anghel,JPA38.9405.2005.Anghel}.

\subsection{BOSONIC SINGLE PARTICLE ENERGIES
\label{BoseDOS}}

\begin{figure}[t]
\begin{center}
\unitlength1mm\begin{picture}(80,65)(0,0)
\put(0,0){\psfig{file=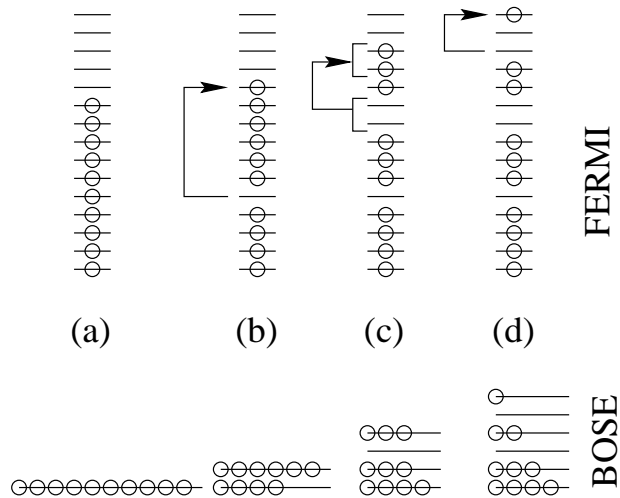,width=80mm}}
\end{picture}
\caption{The correspondence between the Fermi and Bose microscopic 
distributions: (a) both systems are on the ground state; (b) four particles 
on the ground state, $\ebo$, and six particles on the first excited 
state, $\epsilon_{\rm B,1}$; 
(c) four particles on $\ebo$, three on $\epsilon_{\rm B,1}$, and 
three on $\epsilon_{\rm B,3}$ ($\epsilon_{\rm B,2}$ is free); 
(d) final distribution, with four particles on $\ebo$, three on 
$\epsilon_{\rm B,1}$, zero on $\epsilon_{\rm B,2}$, two on 
$\epsilon_{\rm B,3}$, zero on $\epsilon_{\rm B,4}$, and one on 
$\epsilon_{\rm B,5}$.}
\label{excitations}
\end{center}
\end{figure}

Let's now calculate the bosonic energy levels. 
We shall use $\epsilon$ and $\eb$ to denote the single particle energies 
in the Fermi and Bose systems, respectively. 
If $m$ is the number of particles in the system, the particle number 
dependent Fermi energy is denoted by $\ef(m)$. 
Moreover, in both (Bose and Fermi) systems we number the single particle 
states form zero to infinity, starting from the lowest energy level and going 
up-wards; the order in which we number the degenerate levels is not 
important.
At zero temperature, the fermions occupy all the $N$ states, from zero to 
the Fermi energy, $\ef(N)$. This is the ground state (g.s.) of the Fermi 
system and is put into correspondence with the ground state of the Bose system, 
$\ebo$, like in figure \ref{excitations} (a). 

The correspondence between 
the Fermi and the Bose energy levels is done recursively, starting from 
the lowest energies, as it is depicted in Figs. \ref{excitations} 
(b), (c), and (d). If the lowest hole in the Fermi system is created by 
lifting a particle from the level $i$, of energy 
$\epsilon_i$($<\ef(N)$) to the level $N+1$, then in the Bose system 
$i-1$ particles are left on the g.s. and $N-i+1$ particles are lifted 
on the first excited level, $\epsilon_{\rm B,1}$ 
[figure \ref{excitations} (b)]. 
The excitation energy in both systems is the same, so we define 
\[
\epsilon_{\rm B,1} \equiv \frac{\ef(N+1)-\epsilon_i}{N-i+1} \, .
\]
The recursion procedure continues upwards; let's say that we arrived 
at the level $\epsilon_j$. We assume that we lifted 
$k$ particles from the energy levels below $\epsilon_{j}$, so 
we created already $k$ quasiparticle energy levels 
in the Bose system. At least from $\epsilon_{j}$, upwards, 
all the Fermi levels are occupied up to $\ef(N+k)$ (i.e. the 
total number of particles, $N$, plus the $k$ particles that have 
been lifted above the Fermi energy) and all these particles 
stay on the highest occupied energy level in the Bose system, 
$\epsilon_{{\rm B},k}$. Let us now 
lift the $n$ particles from the energy levels $j+1, \ldots, j+n$, onto 
$N+k+1,\ldots,N+k+n$. This corresponds in the 
Bose system to $n-1$ free states, say 
$\epsilon_{{\rm B},k+1}, \ldots,\epsilon_{{\rm B},k+n-1}$, 
and an $n^{\rm th}$ state, $\epsilon_{{\rm B},k+n}$, 
with a population of $N-(j-k)$ particles. 
The difference between the Bose levels which are successively occupied, 
$\epsilon_{{\rm B},k}$ and $\epsilon_{{\rm B},k+n}$, is 
\begin{eqnarray}
\epsilon_{{\rm B},k+n}-\epsilon_{{\rm B},k}  
&\equiv& \frac{\sum_{l=1}^n [\ef(N+k+l)- 
\epsilon_{j+l}]}{N+k-j} 
\approx n\cdot\frac{[\ef(N+k)- \epsilon_{j}]}{N+k-j} 
\label{epsdiff} 
\end{eqnarray} 
(see figure \ref{excitations}). 

The procedure continues upwards until we reach the highest energy particles 
in the Fermi system. 

\subsection{CONTINUOUS LIMIT \label{CL}}

To discuss the thermodynamic limit, we need to go from the discrete case 
to the (quasi)continuous limit. 
For this, we denote the fermionic and the bosonic DOSs by 
$\sigma(\epsilon)$ and $\sib(\eb)$, respectively. 
We assume that in the Fermi system $N_0$ particles are 
condensed \cite{JPA36.L577.2003.Anghel}, so there are no holes in the 
energy interval from 0 to $\ef(N_0)$. Above $\ef(N_0)$ we divide 
the single particle energy axis  into 
microscopic intervals, numbered $i=1,\ldots,\infty$. 
Then the interval $i$ of length $\delta\epsilon_i$, contains 
$d_i=\sigma(\epsilon_i)\delta\epsilon_i$ single particle states. 
The population of the interval $\delta\epsilon_i$ is denoted as 
$\delta n_i\equiv f(\epsilon_i) d_i$ and {\em we assume} that 
$f(\epsilon_i)<1$ for any $i\ge 1$. 
After EST, $N_0$ represents the number of particles on the 
Bose g.s., $\ebo$, and the intervals $\delta\epsilon_i$ transform into 
the intervals $\delta\epsilon_{{\rm B},i}$. The number of bosonic 
states in each of these intervals is 
$d_{{\rm B},i}=d_i-\delta n_i+1\approx d_i-\delta n_i$ (for large enough 
$d_{{\rm B},i}$). Although it has no relevance in the continuous limit, 
note that because of an overlap between successive 
Bose intervals, the expression $d_{{\rm B},i}=d_i-\delta n_i$, leads 
to a better counting of total number of bosonic states (the same argument 
is in \cite{PhysRevB.60.6517.1999.Murthy}). 
Applying equation (\ref{epsdiff}), 
$\delta\epsilon_{{\rm B},i}$ 
is calculated by exciting $[1-f(\epsilon_i)]d_i$ fermions from the 
interval $\delta\epsilon_{i}$, which leads to 
\begin{eqnarray}
\delta\epsilon_{{\rm B},i}&=& (d_i-\delta n_{i})\cdot
\frac{ \epsilon_{\rm F}[N+1+\sum_{j=1}^{i}(d_i-\delta n_{j})]-\epsilon_i}{
N-N_0-\sum_{j=1}^{i}\delta n_{j}} \, . \label{enBmicro}
\end{eqnarray}
From equation (\ref{enBmicro}) we obtain the Bose DOS, 
\begin{eqnarray}
\sigma_{\rm B}(\epsilon_{{\rm B},i}) &=& 
\frac{d_{{\rm B},i}}{\delta\epsilon_{{\rm B},i}} = 
\frac{N-N_0-\sum_{j=1}^{i}\delta n_{j}}{ 
\epsilon_{\rm F}[N+1+\sum_{j=1}^{i}(d_j-\delta n_{j})]-\epsilon_i}\,  
\label{boseDOS}
\end{eqnarray}
and the Bose population (the number of particles divided by the number of 
states), 
\begin{equation} \label{bosepop}
b(\epsilon_{{\rm B},i})\equiv\frac{\delta n_i}{d_i-\delta n_i} 
= \frac{1}{f^{-1}[\epsilon_i(\epsilon_{{\rm B},i})]-1} \, .
\end{equation}
Changing the summations into integrals, equation (\ref{boseDOS}) becomes 
\begin{eqnarray}
\sigma_{\rm B}(\eb) &=&\frac{N-N_0-\int_{0}^{\eb}
b(\eb^\prime)\sib(\eb^\prime)\rmd\eb^\prime}{\epsilon_{\rm F}[N+1+
\int_{0}^{\eb}\sib(\eb^\prime)\rmd\eb^\prime]-\epsilon(\eb)} \, , 
\label{boseDOSi}
\end{eqnarray}
where $\epsilon(\eb)$ is the single particle energy in the original Fermi 
systems that corresponds to the Bose energy $\eb$. 

To calculate $\epsilon(\eb)$, note that the number of Fermi states 
in the small energy interval $\delta\epsilon$ which 
corresponds to the Bose energy interval $\delta\eb$ is equal to the number 
of particles in the interval, plus the number of bosonic energy levels: 
$d(\delta\epsilon)=[b(\eb)+1]\sib(\eb)\delta\eb$. Integrating over all 
the Bose energy levels, we get the expression 
\begin{equation} \label{epsB}
\epsilon(\eb) \equiv \ef\left\{N_0+\int_0^{\eb}\sib(\epsilon)[b(\epsilon)+1] 
\rmd\epsilon \right\} \, .
\end{equation}
The total Bose energy, 
\begin{equation}
E_{\rm B} = \int_0^\infty \rmd\eb\,\sib(\eb)\eb b(\eb)\, , \label{entot}
\end{equation}
is nothing but the excitation energy of the system, in agreement also with 
the definitions given in \cite{JPA35.7255.2002.Anghel}. 

\section{INTERACTION IN THE BOSE SYSTEM \label{BoseInt}}

\begin{figure}[t]
\begin{center}
\unitlength1mm\begin{picture}(45,60)(0,0)
\put(0,0){\psfig{file=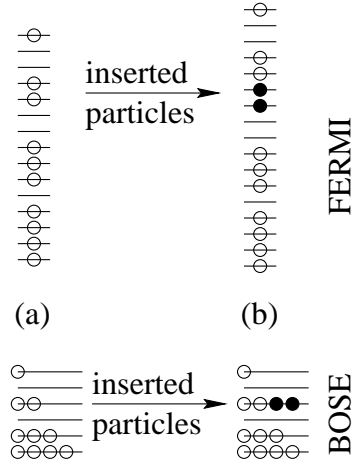,width=45mm}}
\end{picture}
\caption{Example of insertion of two particles (filled circles) on the 
fourth Bose energy level.}
\label{insertion}
\end{center}
\end{figure}

In the general case, the Bose system obtained after the EST is not 
an ideal system. A change in the population of Bose 
``single particle'' states, say $\delta b(\eb)$, changes all the 
single particle energies and, further, $\sib$. To see this, let us 
insert $\xi$ particles in the interval $i$ of the Fermi system. By 
doing this, we also expand the interval by $\xi$ states, so that the 
number of Bose states in any of the intervals would not change 
(see figure \ref{insertion}). 
All the quantities that are affected by this insertion of quasiparticles 
will bear a superscript which refers to the 
energy interval ($i$), or directly to the energy levels [$\ei$ (Fermi) 
or $\ebi$ (Bose)] where the insertion took place. 
Applying equation (\ref{enBmicro}), we calculate the new energy interval 
$\delta\epsilon^i_{{\rm B},k}=\epsilon^i_{{\rm B},k}-\epsilon^i_{{\rm B},k-1}$:
\begin{eqnarray}
\delta\ebki &=& \left\{
\epsilon_{\rm F}[N+1+\xi+\sum_{j=1}^{k}(d_j-\delta n_{j})] -
\ef\left[N_0+\xi+\sum_{j=1}^k d_j\right]\right\} \nonumber \\
&& \cdot \frac{d_k-\delta n_{k}}{N-N_0-\sum_{j=0}^{k}\delta n_{j}} \, ,
\ {\rm if}\ i<k , \label{ilessk}
\end{eqnarray}
or
\begin{eqnarray}
\delta\ebki &=& \left\{
\epsilon_{\rm F}[N+1+\xi+\sum_{j=1}^{k}(d_j-\delta n_{j})] -\ef\left[N_0+
\sum_{j=1}^k d_j\right]\right\} \nonumber \\
&& \times \frac{d_k-\delta n_{k}}{N+\xi-N_0-\sum_{j=0}^{k}\delta n_{j}} \, ,
\ {\rm if}\ i\ge k\, . \label{igek} 
\end{eqnarray}
If $\xi\ll\delta n_i$, then 
\begin{eqnarray}
\delta\ebki &\approx& \delta\ebk+ \xi
\frac{d-\delta n_{k}}{N-N_0-\sum_{j=0}^{k}\delta n_{j}}  
\left\{\frac{\rmd\ef}{\rmd N}[N+1+\sum_{j=1}^{k}(d_j-\delta n_{j})] \right.
\nonumber \\
&& \left.-\frac{\rmd \ef}{\rmd N}[N_0+\sum_{j=1}^{k}d_j]\right\} \, ,
\ {\rm if}\ i<k \label{ilessk0}
\end{eqnarray}
or
\begin{eqnarray}
\delta\ebki&=& \delta\ebk - \xi\frac{\delta\ebk}{N-N_0-
\sum_{j=0}^{k}\delta n_{j}} + 
\xi\frac{d-\delta n_{k}}{N-N_0-\sum_{j=0}^{k}\delta n_{j}} \nonumber \\
&&\times\frac{\rmd \ef}{\rmd N}[N+1+\sum_{j=1}^{k}(d-\delta n_{k})]\, ,
\ {\rm if}\ i\ge k\, . \label{igek0} 
\end{eqnarray}
If in the Bose system we add up the energies of all particles, we obtain 
\[
E_\xi = E_{\rm B} + \xi\ebi + E_{\rm I}\, 
\]
where $E_{\rm B}$ is given by equation (\ref{entot}) and 
$E_{\rm I}\propto\xi$ is the interaction term. 
We may interpret $\epsilon_{\rm B}$ as the quasiparticle energy, and write it as 
\begin{equation} \label{qpen}
\eb = \epsilon_{\rm Bf}+\sum_{\epsilon_{\rm Bf}^\prime}
v(\epsilon_{\rm Bf},\epsilon_{\rm Bf}^\prime)
\delta n(\epsilon_{\rm Bf}^\prime) \, ,
\end{equation}
where $\epsilon_{\rm Bf}$ is the energy of the ``free particles'', 
while $v$ is the interaction potential. 
Note that $v(\epsilon_{\rm Bf},\epsilon_{\rm Bf}^\prime)$ depends on 
$n(\epsilon_{\rm Bf}^\prime)$, for any value of $\epsilon_{\rm Bf}^\prime$. 


\subsection{CALCULATION OF THE INTERACTION POTENTIAL \label{IP}}

The Bose quasiparticle energies and the interaction potential $v$ from 
equation (\ref{qpen}) may be calculated by adding the microscopic 
energy intervals given by the equations (\ref{ilessk0}) 
and (\ref{igek0}). Transforming the summations into integrals and using 
the equations (\ref{boseDOS}), (\ref{boseDOSi}) and (\ref{epsB}) we obtain 
%
%
\begin{eqnarray}
&&\eb^{\ebi}(\xi) = \epsilon_{\rm B} - \xi\int_0^{\eb} 
\frac{\rmd \eb}{N-N_0-\int_{0}^{\eb}
b(\epsilon)\sib(\epsilon)\rmd\epsilon}  \nonumber\\
%
&& + \xi\int_0^{\eb} \frac{\frac{\rmd \ef}{\rmd N}[N+1+
\int_{0}^{\eb}\sib(\epsilon)\rmd\epsilon]\,\rmd \eb}{\ef[N+1+
\int_{0}^{\eb}\sib(\epsilon)\rmd\epsilon]-\ef\left\{N_0+
\int_0^{\eb}\sib(\epsilon)[b(\epsilon)+1]\rmd\epsilon 
\right\}} \nonumber \\
&&= \eb - \xi\int_0^{\eb}\frac{1-\sib(\eb)\frac{\rmd \ef}{\rmd N}[N+1+
\int_{0}^{\eb}\sib(\epsilon)\rmd\epsilon]}{N-N_0-\int_{0}^{\eb}
b(\epsilon)\sib(\epsilon)\rmd\epsilon}\rmd \eb \, ,\label{epsm} \\
&&{\rm for}\ \ebi\ge\eb \nonumber 
\end{eqnarray}
and
\begin{eqnarray}
&&\eb^{\ebi}(\xi) = \eb - \xi\int_0^{\ebi}\frac{1-\sib(\eb)
\frac{\rmd \ef}{\rmd N}[N+1+\int_{0}^{\eb}\sib(\epsilon)\rmd\epsilon]
}{N-N_0-\int_{0}^{\eb}b(\epsilon)\sib(\epsilon)\rmd\epsilon}
\rmd \eb \nonumber \\
&&+ \xi\int_{\ebi}^{\eb}\frac{\rmd \eb\left\{\frac{\rmd \ef}{\rmd N}
[N+1+\int_{0}^{\eb}\sib(\epsilon)\rmd\epsilon] -\frac{\rmd \ef}{\rmd N}[N_0+
\int_0^{\eb}\sib(\epsilon)[b(\epsilon)+1]\rmd\epsilon]
\right\}}{\epsilon_{\rm F}[N+1+\int_{0}^{\eb}\sib(\epsilon)\rmd\epsilon]
-\ef\left\{N_0+\int_0^{\eb}\sib(\epsilon)[b(\epsilon)+1]\rmd\epsilon 
\right\}} \nonumber\\
&&= \eb - \xi\int_0^{\ebi}\frac{1-\sib(\eb)\frac{\rmd \ef}{\rmd N}[N+1+
\int_{0}^{\eb}\sib(\epsilon)\rmd\epsilon]}{N-N_0-\int_{0}^{\eb}
b(\epsilon)\sib(\epsilon)\rmd\epsilon}\rmd \eb 
+ \xi\int_{\ebi}^{\eb}\sib(\eb) \nonumber \\
%
&& \times\frac{\left\{\frac{\rmd \ef}{\rmd N}[N+1+
\int_{0}^{\eb}\sib(\epsilon)\rmd\epsilon] - \frac{\rmd \ef}{\rmd N}[N_0+
\int_0^{\eb}\sib(\epsilon)[b(\epsilon)+1]\rmd\epsilon]
\right\}}{N-N_0-\int_{0}^{\eb}b(\epsilon)\sib(\epsilon)\rmd\epsilon}
\,\rmd\eb \nonumber\\ 
&&{\rm for}\  \ebi<\eb . \label{epsM}
\end{eqnarray}
In terms of $\epsilon$, and using the shorthand notation 
$\epsilon_0\equiv\ef(N_0)$, the equations (\ref{epsm}) and 
(\ref{epsM}) become 
\begin{eqnarray}
&&\eb^{\ei}(\xi) = \eb - \xi\int_{\epsilon_0}^{\epsilon(\eb)}
\rmd\epsilon_1\, 
\frac{\sigma(\epsilon_1)[1-f(\epsilon_1)]}{\sib(\eb(\epsilon_1))} 
\left\{\left[N-N_0-\int_{\epsilon_0}^{\epsilon_1}\rmd\epsilon_2\,
\sigma(\epsilon_2) f(\epsilon_2)\right]^{-1} \right. \nonumber \\
%
&&- \left.\frac{\sigma^{-1}\left[\ef\left(N+\int_{\epsilon_0}^{\epsilon_1}
\rmd\epsilon_2\,\sigma(\epsilon_2)(1-f(\epsilon_2))\right)\right]}{
\ef\left[N+\int_{\epsilon_0}^{\epsilon_1}\rmd\epsilon_2\,\sigma(\epsilon_2)
(1-f(\epsilon_2))\right]-\epsilon_1} \right\} \, ,\ {\rm for}\ 
\ei\ge\epsilon(\eb)
\label{epsm1}
\end{eqnarray}
and
\begin{eqnarray}
&& \eb^{\ei}(\xi) = \eb -\xi\int_{\epsilon_0}^{\ei}\rmd\epsilon_1\, 
\frac{\sigma(\epsilon_1)[1-f(\epsilon_1)]}{\sib(\eb(\epsilon_1))} 
\left\{\left[N-N_0-\int_{\epsilon_0}^{\epsilon_1}\rmd\epsilon_2\,
\sigma(\epsilon_2) f(\epsilon_2)\right]^{-1} \right. \nonumber \\
&&- \left. 
\frac{\sigma^{-1}\left[\ef\left(N+\int_{\epsilon_0}^{\epsilon_1}
\rmd\epsilon_2\,\sigma(\epsilon_2)(1-f(\epsilon_2))\right)\right]}{
\ef\left[N+\int_{\epsilon_0}^{\epsilon_1}\rmd\epsilon_2\,\sigma(\epsilon_2)
(1-f(\epsilon_2))\right]-\epsilon_1} \right\}
+ \xi\int_{\ei}^{\epsilon(\eb)}\rmd\epsilon_1\, 
\sigma(\epsilon_1)[1-f(\epsilon_1)]
\nonumber \\
&&\times
\frac{\sigma^{-1}\left[\ef\left(N+\int_{\epsilon_0}^{\epsilon_1}
\rmd\epsilon_2\,\sigma(\epsilon_2)(1-f(\epsilon_2))\right)\right] - 
\sigma^{-1}(\epsilon_1)}{N-N_0-\int_{\epsilon_0}^{\epsilon_1}\rmd\epsilon_2\,
\sigma(\epsilon_2) f(\epsilon_2)} \, ,\ {\rm for}\ 
\ei<\epsilon(\eb) , \label{epsM1}
\end{eqnarray}
respectively. 
The terms proportional to $\xi$ in equations (\ref{epsm}), (\ref{epsM}), 
(\ref{epsm1}), and (\ref{epsM1}) represent the interaction term 
$v(\eb,\ebi)$ of equation (\ref{qpen}). 

\subsection{BOSE DENSITY OF STATES}

The changes of the single particle energy levels produce changes of 
the Bose DOS, $\sib(\eb)$. 
The new DOS is obtained by using eqs. (\ref{ilessk0}) and (\ref{igek0}):
\begin{eqnarray}
&&\sib^i(\epsilon_{{\rm B},k})=\frac{d-\delta n_{k}}{
\delta\epsilon^i_{{\rm B},k}} \approx 
\sib(\epsilon_{{\rm B},k}) \cdot\left(1-\xi
\frac{\sib(\epsilon_{{\rm B},k})}{N-N_0-\sum_{j=1}^{k}\delta n_{j}}
\right. \label{varsiga} \\
&&\times \left. 
\left\{\frac{\rmd \ef}{\rmd N}[N+1+\sum_{j=1}^{k}(d_j-\delta n_{j})]-
\frac{\rmd \ef}{\rmd N}[N_0+\sum_{j=1}^{k}d_j] \right\} \right) \, ,
\ {\rm if}\ i<k, \nonumber 
\end{eqnarray}
and
\begin{eqnarray}
&&\sib^i(\epsilon_{{\rm B},k})=\frac{d-\delta n_{k}}{
\delta\epsilon^i_{{\rm B},k}} \approx 
\sib(\epsilon_{{\rm B},k}) \cdot\left(1+\frac{\xi}{N-N_0-
\sum_{j=0}^{k}\delta n_{j}} \right. \label{varsigb} \\
%
&&-\left.\xi\frac{\sib(\epsilon_{{\rm B},k})}{N-N_0-\sum_{j=1}^{k}
\delta n_{j}}\frac{\rmd \ef}{\rmd N}[N+1+\sum_{j=1}^{k}(d-\delta n_{j})]
\right)\, ,\ {\rm if}\ i\ge k\, . \nonumber
\end{eqnarray}

Writing equations (\ref{epsm}), (\ref{epsM}), (\ref{epsm1}) and 
(\ref{epsM1}) in the format 
\begin{equation} \label{vareps}
\ebki=\ebk+\xi^{\ebi}\cdot\frac{\partial\ebk}{\partial\xi^{\ebi}} \, ,
\end{equation}
and using equations (\ref{varsiga}) and (\ref{varsigb}) we obtain
\begin{eqnarray}
&&\sibeps(\epsilon^0_{{\rm B},k}+\xi^{\ebi}
\cdot\frac{\partial\ebk}{\partial\xi^{\ebi}})= 
\sib(\ebk) \cdot\left(1-\frac{\xi^{\ebi}\sib(\ebk)}{
N-N_0-\int_{0}^{\ebk}b(\ebp)\sib(\ebp)\rmd\ebp}
\right. \nonumber \\
&&\times \left. \left\{\left.\frac{\rmd \ef}{\rmd N}\right|_{\left[
N+1+\int_{0}^{\ebk}\sib(\ebp)\rmd\ebp\right]}
-\left.\frac{\rmd \ef}{\rmd N}\right|_{\left\{N_0+\int_0^{\eb}
\sib(\ebp)[b(\ebp)+1]\rmd\ebp \right\}} \right\} \right) \, ,
\nonumber \\
&& \ {\rm if}\ \ebi<\ebk, \label{varsig1} 
\end{eqnarray}
and 
\begin{eqnarray}
&&\sibeps(\ebk+\xi^{\ebi}\cdot\frac{\partial\ebk}{\partial\xi^{\ebi}})
= \sib(\ebk) \cdot\left(1+
\frac{\xi^{\ebi}}{N-N_0-\int_{0}^{\ebk}b(\ebp)\sib(\ebp)\rmd\ebp} 
\right. \nonumber\\
%
&&\times\left.\left\{1-\sib(\ebk)\left.
\frac{\rmd \ef}{\rmd N}\right|_{\left[N+1+\int_{0}^{\ebk}
\sib(\ebp)\rmd\ebp\right]}\right\}\right)\, , \ {\rm if}\ \ebi\ge\ebk\, . 
\label{varsig2} 
\end{eqnarray}
Writing the integrals in terms of $\epsilon$, Eqs. (\ref{varsig1}) 
and (\ref{varsig2}) become
\begin{eqnarray}
&&\sibeps(\ebk+\xi^{\ebi}\cdot\frac{\partial\ebk}{\partial\xi^{\ebi}})
=  \sib(\ebk)\cdot\left(1-\frac{\xi^{\ebi}\sib(\ebk)}{
N-N_0-\int_{\epsilon_0}^{\epsilon_k}\rmd\epsilon_1\,
\sigma(\epsilon_1)f(\epsilon_1)}\right. \nonumber \\
%
&& \times \left. 
\left\{\sigma^{-1}\left[\ef\left(N+\int_{\epsilon_0}^{\epsilon_k}
\rmd\epsilon_1\,\sigma(\epsilon_1)(1-f(\epsilon_1))\right)\right] - 
\sigma^{-1}(\epsilon_k)\right\} \right) \nonumber \\ 
&&\equiv \sib(\ebk)\cdot\left\{1-\xi^{\ebi} \cdot
\frac{\sigma^{-1}\left[\ef\left(N+\int_{\epsilon_0}^{\epsilon_k}
\rmd\epsilon_1\,\sigma(\epsilon_1)(1-f(\epsilon_1))\right)\right] - 
\sigma^{-1}(\epsilon_k)}{\ef\left[N+\int_{\epsilon_0}^{\epsilon_k}
\rmd\epsilon_1\,\sigma(\epsilon_1)(1-f(\epsilon_1))\right] - 
\epsilon_k} \right\} \, , \nonumber \\
&&\ {\rm if}\ \ebi<\ebk \label{varsig1eps}
\end{eqnarray}
and 
\begin{eqnarray}
&&\sibeps(\ebk+\xi^{\ebi}\cdot\frac{\partial\ebk}{\partial\xi^{\ebi}})
= \sib(\ebk) \cdot\left(1+\frac{\xi^{\ebi}}{N-N_0-
\int_{\epsilon_0}^{\epsilon_k}\rmd\epsilon_1\,
\sigma(\epsilon_1)f(\epsilon_1)} \right. \nonumber \\
%
&&\times\left.\left\{1-\frac{\sib(\ebk)}{
\sigma\left[\ef\left(N+\int_{\epsilon_0}^{\epsilon_k}
\rmd\epsilon_1\,\sigma(\epsilon_1)(1-f(\epsilon_1))\right)\right]}
\right\}\right)\, , \ {\rm if}\ \ebi\ge\ebk\, . \label{varsig2eps}
\end{eqnarray}
%


\section{FINDING THE EQUILIBRIUM DISTRIBUTION \label{eq_distr}}

The Bose and the Fermi systems are thermodynamically equivalent by 
construction. Therefore, the equilibrium particle distribution in the Fermi 
gas, if transformed by (\ref{bosepop}), should give us the equilibrium 
particle distribution in the Bose gas and knowing this, we can calculate 
all the microscopic parameters of the Bose system and the 
quasiparticle-quasiparticle interactions like in sections \ref{CL} and 
\ref{IP}, determining in this way all the properties of the gas. 
The Fermi equilibrium (i.e. most probable) particle distribution is the 
well known $f(\epsilon)=[\exp(\beta\epsilon-\beta\mu)+1]^{-1}$ (where 
$\beta\equiv(\kb T)^{-1}$), and this gives us by (\ref{bosepop}), the 
bosonic population 
\begin{equation}\label{bosepopt}
b(\eb)=\exp[\beta\mu-\beta\epsilon(\eb)] \, .
\end{equation}
which can be readily used in the calculations. 

Nevertheless, we have still another method to calculate the equilibrium 
particle distribution in the Bose gas. In sections \ref{iEST} and 
\ref{BoseInt} we defined self-consistently the quasiparticle energy 
levels in the Bose gas and the quasiparticle-quasiparticle interaction, 
so we can use this as a starting point to build the Bose entropy, $\SB$, 
and canonical partition function, $\ZB$. Then the equilibrium particle 
distribution in the Bose gas should be the one that maximizes $\ZB$ 
at constant $N$ and $\UB$. The Bose and the Fermi 
systems are thermodynamically equivalent, so the canonical partition 
function of the Fermi system, $\ZF$, is identical to $\ZB$ by construction, 
and indeed, the distribution that maximizes one should correspond (by an 
EST transformation \ref{bosepop}) to the distribution that maximized the 
other, unless the canonical and grandcanonical ensembles in at least 
one of the systems are not equivalent. 

To check this equivalence, let me calculate $\SB$ and maximize it 
at constant $\UB$ and $N$. (Note that constant $\UB$ and $N$ is equivalent 
to constant $U$ and $N$.) So let again $\delta\ebk$ be the 
microscopic energy intervals along the $\eb$ axis, each containing 
$\dbk$ states. The total number of configurations available in one 
interval is 
\begin{equation} \label{wbk}
\wbk = \frac{(\delta n_k+\dbk)!}{\delta n_k! \dbk!} \, . 
\end{equation}
Summing up 
the contributions of all the intervals and applying the Stirling 
approximation, $\log m!\approx m\log(m/\e)$, we obtain the Bose entropy:
\begin{equation} 
S_{\rm B}/\kb = \int_{0}^\infty\rmd\eb\, \sib(\eb)\{[1+b(\eb)]
\log[1+b(\eb)] - b(\eb)\log[b(\eb)]\} \, , \label{boseS}
\end{equation}
To find the equilibrium configuration, we have to maximize $S_{\rm B}$ 
as a functional of $b$, at fixed $E_{\rm B}$ and $N$. For this we 
introduce the Lagrange multipliers $\beta_{\rm B}$ and $\beta_{\rm B}\mub$, 
corresponding to $E_{\rm B}$ and $N$, respectively, and we solve 
\begin{eqnarray}
\frac{\delta(S/\kb-\beta_{\rm B}E_{\rm B}+\beta_{\rm B}\mub N)}{\delta b}
&=& 0 \, , 
\label{bosepopB}
\end{eqnarray}
where, as shown in section \ref{iEST}, from thermodynamic equivalence 
arguments we have $\beta_{\rm B}\equiv\beta$ and 
$\mub\equiv\mu-\ef$ (see also Ref. \cite{JPA35.7255.2002.Anghel}). 
The functional derivative $\delta E_{\rm B}/\delta b(\eb)$ is 
not simply $\eb$. To calculate it we rewrite Eq. (\ref{qpen}) as 
\begin{equation} \label{qpenint}
\eb^{\delta b} = \eb+ \int_0^\infty \rmd\eb^\prime\, 
v(\eb,\eb^\prime) \sib(\eb^\prime)\delta b(\eb^\prime) \, ,
\end{equation}
where $\eb$ is the quasiparticle energy at equilibrium configuration and 
$\delta b(\eb)$ is the deviation of the population from the 
equilibrium configuration, $b(\eb)$. Note that both, $b(\eb)$ and 
$\delta b(\eb)$ are given as functions of the energy before the 
insertion of particles, $\eb$. 
The interaction term is calculated in Section \ref{IP}. 
Using (\ref{qpenint}) we calculate the energy 
\begin{eqnarray}
E_{\rm B}^\prime &=& \int_0^\infty \rmd\eb^{\delta b}\, \eb^{\delta b}
[\sib(\eb^{\delta b})+\delta\sib(\eb^{\delta b})][b(\eb)+
\delta b(\eb)] \label{enB} \\
&=& \int_0^\infty \rmd\eb\, \left[ \eb+ \int_0^\infty \rmd\eb^\prime\, 
v(\eb,\eb^\prime) \sib(\eb^\prime)\delta b(\eb^\prime)\right] 
\sib(\eb)[b(\eb)+\delta b(\eb)] \, , \nonumber 
\end{eqnarray}
where we used the equation $\rmd\eb^{\delta b}/\rmd\eb=\sib(\eb)/[\sib(\eb^{\delta b})+\delta\sib(\eb^{\delta b})]$ and the obvious notation 
$b(\eb)\equiv b[\eb(\eb^{\delta b})]$ in the first line. 
From (\ref{enB}), and keeping only the first order in $\delta b$, we 
obtain 
\begin{eqnarray}
&&E_{\rm B}^\prime = E_{\rm B} 
+ \int_0^\infty \rmd\eb\, \eb \sib(\eb)\delta b(\eb) \label{enB1} \\
&&+ \int_0^\infty \rmd\eb\int_0^\infty \rmd\eb^\prime\, b(\eb)\sib(\eb)
v(\eb,\eb^\prime) \sib(\eb^\prime)\delta b(\eb^\prime) \, . \nonumber
\end{eqnarray}
The first integral in equation (\ref{enB1}), let's call it $E_{\rm qp}$, 
is the typical contribution 
of the added quasiparticle energies to the total energy of the system, 
whereas the double integral (say $E_\mu$) is due to the collective change 
of quasiparticle energies and can not be eliminated by a redefinition of the 
ground state energy, as e.g. in the Fermi liquid theory. 
So $E_\mu$ will produce an effective quasiparticle energy 
\begin{equation}
\bar\eb(\eb) = \eb+\int_0^\infty \rmd\eb^\prime\, b(\eb^\prime)\sib(\eb^\prime)
v(\eb^\prime,\eb) \equiv \eb+\tilde\eb(\eb) \,. \label{bareb}
\end{equation}
Since $\delta S/\delta b|_{\eb}=\kb\sib(\eb)\log[1+b^{-1}(\eb)]$ and 
$\delta N/\delta b|_{\eb}=\sib(\eb)$, plugging (\ref{enB1}) into 
equation (\ref{bosepopB}), one obtains the following equation for 
the equilibrium distribution: 
\begin{equation} \label{bosepopB1}
\log[1+b^{-1}(\eb)] - \beta\bar\eb(\eb) +\beta\mub  = 0 \, .
\end{equation}
Equation (\ref{bosepopB1}) leads to the typical Bose quasiparticle 
level population, 
\begin{equation} \label{bosedistr}
b(\eb)=[\exp(\beta\bar\eb(\eb)-\beta\mub)-1]^{-1} \, . 
\end{equation}

If neither the Fermi, not the resultant Bose system is pathologic, then 
the expressions (\ref{bosepopt}) and (\ref{bosedistr}) should give the 
same result, eventually with some corrections which vanish in the 
thermodynamic limit. 


\section{COMPARISON OF THE BOSE AND FERMI DESCRIPTIONS \label{test}}

To make concrete calculations and check if Eqs. (\ref{bosepopt}) and 
(\ref{bosedistr}) give the same results, we assume that the Fermi DOS has 
the general form $\sigma(\epsilon)=C\epsilon^s$, where $C$ and $s$ are 
constants and we calculate the Bose 
energies, DOS, and chemical potential by using the Fermi distribution, 
$f(\epsilon)=[\exp(\beta\epsilon-\beta\mu)+1]^{-1}$, in the formulas of 
Section \ref{CL}. Using these results and the expressions for 
$v(\eb^\prime,\eb)$ calculated in Section \ref{BoseInt}, we can 
compare Eqs. (\ref{bosepopt}) and (\ref{bosedistr}). 

\subsection{CONSTANT DENSITY OF STATES \label{cDOS}}

In Ref. \cite{JPA35.7255.2002.Anghel} it is proven that the Bose and Fermi 
descriptions (obtained by EST) of systems with constant DOS are 
equivalent under canonical 
conditions. Therefore we start by using such systems to test the 
formalism introduced above.
If the density of states is constant ($\sigma\equiv C$) then 
$\ef(n) = n/C$ and by using (\ref{epsB}) to rewrite 
$\epsilon_i$ of equation (\ref{boseDOS}) as 
\[
\epsilon_i=C^{-1}\cdot\left[N_0+1+\sum_{j=1}^{i}d_i\right] \, ,
\]
the Bose DOS becomes \cite{JPA35.7255.2002.Anghel}
\begin{equation}
\sigma_{\rm B}(\epsilon_{{\rm B},i}) \equiv
C = \sigma \, . \label{boseDOS2D}
\end{equation}
Moreover, equation (\ref{enBmicro}) becomes 
\begin{equation} \label{enBmicro2D}
\delta\epsilon_{{\rm B},i} = C^{-1}\cdot(d_i-\delta n_{i}) \, ,
\end{equation}
which, in the continuous limit leads to the Bose energies 
\begin{equation}
\eb=\int_0^\epsilon[1-f(\epsilon_1)]\rmd\epsilon_1=\kb T\log\left[
\frac{1+\e^{\beta(\epsilon-\mu)}}{1+\e^{-\beta\mu}}\right]
\, . \label{epsBeps2D}
\end{equation}
From equations (\ref{ilessk0}) and (\ref{igek0}) follows that 
$v\equiv 0$, so the Bose gas is also noninteracting, with 
$\mub=\mu-\ef$ \cite{JPA35.7255.2002.Anghel} and $\tilde\eb=0$.

To test the Bose-Fermi description equivalence we invert equation 
(\ref{epsBeps2D}) to get 
\begin{equation}
\e^{\beta(\epsilon-\mu)}=\e^{\beta\eb}
\left[1+\e^{-\beta\mu}\right] - 1 \, , \label{epsFepsB2D}
\end{equation}
which, if plugged into (\ref{bosepopt}), gives 
\begin{equation} \label{b02D}
b=\left[\e^{\beta\eb}\left[1+\e^{-\beta\mu}\right]-1\right]^{-1} .
\end{equation}
But for a Fermi gas of constant DOS we have the identity 
(from equation 4, Ref. \cite{JPA35.7255.2002.Anghel}) 
\begin{equation}
\exp(\beta\mu)=\exp(\beta\ef)-1 \, , \label{muef2D}
\end{equation}
so 
\begin{equation} \label{mub2D}
1+\e^{-\beta\mu} = 1+[\exp(\beta\ef)-1]^{-1}=\exp[\beta(\ef-\mu)] 
\equiv \exp[-\beta\mub] \, .
\end{equation}
Plugging (\ref{epsFepsB2D}) into (\ref{mub2D}) we obtain equation 
(\ref{bosedistr}), 
\begin{equation} \label{equiv2D}
b=\exp[\beta(\epsilon-\mu)]=
\left(\exp\{\beta[\eb(\epsilon)-\mub]\}-1\right)^{-1} \, ,
\end{equation}
which proves the equivalence between the Bose and Fermi descriptions of 
the gas with constant DOS.

The equivalence of the Bose and Fermi descriptions 
does not prove the canonical-grandcanonical ensemble equivalence. The 
ensemble inequivalence in condensed Bose systems have been extensively 
studied (see \cite{AnnPhysNY270.198.1998.Holthaus,AnnPhysNY276.321.1999.Holthaus} and references therein); from the EST perspective, the fluctuations 
of the Bose and Fermi energy level populations for systems of constant DOS have 
been compared in Ref. \cite{JPA38.9405.2005.Anghel}. 

\subsection{NON-CONSTANT DENSITY OF STATES \label{ncDOS}}

Again, if we take $f(\epsilon)=[\exp(\beta\epsilon-\beta\mu)+1]^{-1}$, 
then $N_0=0$ (no ``condensation''). 
If $\sigma(\epsilon)\equiv C\epsilon^s$ with $s\ne0$, then 
\begin{equation}
N=\int_0^{\ef} C\epsilon^s\,\rmd\epsilon = \frac{C\ef^{s+1}}{s+1} \, ,
\label{eF} 
\end{equation}
and 
\begin{equation}
\sib(\eb(\epsilon))=\frac{N-\int_0^\epsilon
\frac{\sigma(\epsilon)\, \rmd\epsilon}{\e^{\beta(\epsilon-\mu)}+1}}{
\left[\frac{s+1}{C}\right]^{1/(s+1)}\cdot 
\left[N+\int_0^\epsilon\frac{\sigma(\epsilon)\,\rmd\epsilon}{
1+\e^{-\beta(\epsilon-\mu)}}\right]^{1/(s+1)}-\epsilon} \, , \label{sigmaB} 
\end{equation}
where again 
\begin{eqnarray}
b(\eb)&=& \e^{-\beta[\epsilon(\eb)-\mu]} \, . \label{b0} 
\end{eqnarray}
Using the equations (\ref{eF}-\ref{b0}) we rewrite the equation 
(\ref{enBmicro}) as 
%
%
\begin{equation}
\rmd\eb=\rmd\epsilon\cdot\frac{C\epsilon^s}{1+\e^{-\beta(\epsilon-\mu)}} 
\cdot\frac{\left[\frac{s+1}{C}\right]^{1/(s+1)}\cdot 
\left[N+\int_0^\epsilon\frac{\sigma(\epsilon)\,\rmd\epsilon}{
1+\e^{-\beta(\epsilon-\mu)}}\right]^{1/(s+1)}-\epsilon}{
N-\int_0^\epsilon\frac{\sigma(\epsilon)\, \rmd\epsilon}{
\e^{\beta(\epsilon-\mu)}+1}} \, , \label{debdeps} 
\end{equation}
to obtain 
\begin{equation}
\eb = \int_0^\epsilon\left\{\frac{\rmd\epsilon_1\,C\epsilon_1^s}{1+\e^{-\beta(\epsilon_1-\mu)}} 
\cdot\frac{\left[\frac{s+1}{C}\right]^{1/(s+1)}\cdot 
\left[N+\int_0^{\epsilon_1}\frac{\sigma(\epsilon_2)\,\rmd\epsilon_2}{
1+\e^{-\beta(\epsilon_2-\mu)}}\right]^{1/(s+1)}-\epsilon_1}{
N-\int_0^{\epsilon_1}\frac{\sigma(\epsilon_2)\, \rmd\epsilon_2}{
\e^{\beta(\epsilon_2-\mu)}+1}} \right\} . \label{epsBeps}
\end{equation}
For the two cases, $\epsilon^\prime\ge\epsilon$ and 
$\epsilon^\prime<\epsilon$, the interaction potential $v$ is 
\begin{eqnarray}
&&v(\epsilon,\epsilon^\prime\ge\epsilon) = - \int_0^{\epsilon}\rmd\epsilon_1\, 
\frac{1}{\sib(\eb(\epsilon_1))}\frac{C\epsilon_1^s}{1+
\e^{-\beta(\epsilon_1-\mu)}} \nonumber \\ 
&&\times\left\{\left[N-\int_0^{\epsilon_1}
\rmd\epsilon_2\,\frac{C\epsilon_2^s}{1+\e^{\beta(\epsilon_2-\mu)}}
\right]^{-1} - C^{-1}\left(\ef^{s+1}\left[N \right.\right.\right.\nonumber \\ 
&&\left.\left.\left.+\int_0^{\epsilon_1}\frac{C\epsilon_2^s\,\rmd\epsilon_2}
{1+\e^{-\beta(\epsilon_2-\mu)}}\right]-\epsilon_1\ef^s\left[N+\int_0^{\epsilon_1}
\frac{C\epsilon_2^s\,\rmd\epsilon_2}{1+\e^{-\beta(\epsilon_2-\mu)}}\right]
\right)^{-1} \right\} \nonumber \\
&&= - \int_0^{\epsilon}\rmd\epsilon_1\, 
\frac{1}{\sib(\eb(\epsilon_1))}\frac{C\epsilon_1^s}{1+
\e^{-\beta(\epsilon_1-\mu)}} \left\{\left[N-\int_0^{\epsilon_1}
\rmd\epsilon_2\,\frac{C\epsilon_2^s}{1+\e^{\beta(\epsilon_2-\mu)}}
\right]^{-1} \right. \nonumber \\
&&- \left.\left((s+1)\left[N+\int_0^{\epsilon_1}
\rmd\epsilon_2\,\frac{C\epsilon_2^s}{1+\e^{-\beta(\epsilon_2-\mu)}}\right]
-\epsilon_1C^{\frac{1}{s+1}}(s+1)^{\frac{s}{s+1}}
\right. \right. \nonumber \\
&&\times\left. \left. 
\left[N+\int_0^{\epsilon_1}\rmd\epsilon_2\,\frac{C\epsilon_2^s}{1+
\e^{-\beta(\epsilon_2-\mu)}}\right]^{\frac{s}{s+1}}
\right)^{-1} \right\}  \label{ebepsm}
\end{eqnarray}
and
\begin{eqnarray}
&&v(\epsilon,\epsilon^\prime<\epsilon) = 
- \int_0^{\epsilon^\prime}\rmd\epsilon_1\, 
\frac{1}{\sib(\eb(\epsilon_1))}\frac{C\epsilon_1^s}{1+
\e^{-\beta(\epsilon_1-\mu)}} \label{ebepsM} \\ 
&& \times\left\{\left[N-\int_0^{\epsilon_1}
\rmd\epsilon_2\,\frac{C\epsilon_2^s}{1+\e^{\beta(\epsilon_2-\mu)}}
\right]^{-1} \left((s+1)\left[N+\int_0^{\epsilon_1}
\rmd\epsilon_2\,\frac{C\epsilon_2^s}{1+\e^{-\beta(\epsilon_2-\mu)}}\right]
\right.\right. \nonumber\\
&&\left.\left.-\epsilon_1C^{\frac{s}{s+1}}(s+1)^{\frac{s}{s+1}}
\left[N+\int_0^{\epsilon_1}\rmd\epsilon_2\,\frac{C\epsilon_2^s}{1+
\e^{-\beta(\epsilon_2-\mu)}}\right]^{\frac{s}{s+1}}
\right)^{-1} \right\} \nonumber \\
&&+\int_{\epsilon^\prime}^{\epsilon(\eb)}\rmd\epsilon_1\, 
\frac{C\epsilon_1^s}{1+\e^{-\beta(\epsilon_1-\mu)}} \nonumber \\
&&\times
\frac{\left[C^{\frac{1}{s+1}}(s+1)^{\frac{s}{s+1}}\left(N+\int_0^{\epsilon_1}
\rmd\epsilon_2\,\sigma(\epsilon_2)/[1+\e^{-\beta(\epsilon_2-\mu)}]
\right)^{\frac{s}{s+1}}
\right]^{-1} - (C\epsilon_1^s)^{-1}}{N-\int_0^{\epsilon_1}\rmd\epsilon_2\,
\frac{C\epsilon_2}{1+\e^{\beta(\epsilon_2-\mu)}}} \, . \nonumber 
\end{eqnarray}

\subsubsection{Dimensionless expressions} \label{diless}

To identify the relevant parameters of the gas, and for the convenience of 
the calculations, let us isolate dimensionless 
quantities from the equations (\ref{eF}-\ref{epsBeps}). 
Hereafter we shall use extensively the notations $x\equiv\beta\epsilon$, 
$x^\prime\equiv\beta\epsilon^\prime$, $y\equiv\beta\mu$, and also 
their Bose correspondents 
$\xb\equiv\beta\eb$ and $\xbp\equiv\beta\eb^\prime$. 
A quantity that appears in almost all the expressions is the integral 
over Fermi distribution: 
\begin{eqnarray}
Fi(T,C,\epsilon,\mu,s)&\equiv& 
\int_0^\epsilon\frac{C\epsilon_1^{s}\,\rmd\epsilon_1}{
\e^{\beta(\epsilon_1-\mu)}+1} 
= C(\kb T)^{s+1}\int_0^{x=\beta\epsilon}\frac{x_1^{s}\,\rmd x_1}{\e^{x_1-y}+1} 
\nonumber \\
&\equiv& C(\kb T)^{s+1} Fir(x,y,s) \, ,  \label{fir} 
\end{eqnarray}
related to the integral over the hole density, 
\begin{eqnarray}
Fi^\prime(T,C,\epsilon,\mu,s)&\equiv& 
\int_0^\epsilon\frac{C\epsilon_1^{s}\,\rmd\epsilon_1}{
\e^{-\beta(\epsilon_1-\mu)}+1} \equiv C(\kb T)^{s+1} Fir^\prime(x,y,s)
\nonumber \\
&=& C(\kb T)^{s+1}
\left[\frac{x^{s+1}}{s+1}- Fir(x,y,s)\right]\, .  \label{firp}
\end{eqnarray}
Both, $Fi$ and $Fi^\prime$ are defined for $s>-1$. 
Another quantity of interest is the ``reduced'' particle number
\begin{equation} \label{nr}
N_r\equiv \frac{(s+1)N}{C(\kb T)^{s+1}} = \left(\frac{\ef}{\kb T}\right)^{s+1}
\end{equation}
With the definitions (\ref{fir}-\ref{nr}), the density of states becomes 
\begin{eqnarray}
\sib(\eb(\epsilon))&=& 
\frac{C(\kb T)^s}{(s+1)}\frac{\left[N_r-(s+1)Fir(x,y,s)\right]}{
\left[N_r+(s+1)Fir^\prime(x,y,s)\right]^{1/(s+1)}-x} \nonumber \\
&\equiv& \frac{C(\kb T)^s}{(s+1)} \sigma_{{\rm B},r}(N_r(y,s),x,y,s) \, ,
\end{eqnarray}
and the Bose energy may be written as 
\[
\eb=(s+1)\kb T \int_0^x \frac{\rmd x_1\,x_1^s}{1+
\e^{-x_1+y}} \cdot \sigma^{-1}_{{\rm B},r}(N_r(y,s),x_1,y,s) 
\equiv (\kb T) \epsilon_{{\rm B}r}(x,y,s) \, .
\]
The interaction term $v(\eb,\eb^\prime)$ (\ref{qpen}) has two 
different expressions, depending on the sign of $\eb-\eb^\prime$, or 
equivalently, $\epsilon-\epsilon^\prime$: 
\begin{eqnarray}
&&v(\epsilon,\epsilon^\prime\ge\epsilon) = 
- \frac{s+1}{C(\kb T)^s}\int_0^{x} 
\frac{\rmd x_1\,x_1^s}{(1+\e^{-x_1+y})\sigma_{{\rm B}r}} \left\{\frac{s+1}{N_r-
(s+1)Fir(x_1,y,s)} \right. \nonumber \\
&&-\left.\frac{1}{N_r+(s+1)Fir^\prime(x_1,y,s+1)- x_1
\left[N_r+(s+1)Fir^\prime(x_1,y,s)\right]^{s/(s+1)}}\right\} 
\nonumber \\ && \label{varepsxp1}
\end{eqnarray}
and
\begin{eqnarray}
&&v(\epsilon,\epsilon^\prime<\epsilon) = -\frac{s+1}{C(\kb T)^s}
\int_0^{x^\prime}\frac{\rmd x_1\, x_1^s}{(1+\e^{-x_1+y})\sigma_{{\rm B}r}} 
\left\{\frac{s+1}{N_r-(s+1)Fir(x_1,y,s)} \right. \nonumber \\
&&-\left.\frac{1}{N_r+(s+1)Fir^\prime(x_1,y,s)- x_1
\left[N_r+(s+1)Fir^\prime(x_1,y,s)\right]^{s/(s+1)}}\right\} \nonumber \\
&&+ \frac{s+1}{C(\kb T)^s}\int_{x^\prime}^x\rmd x_1\, 
\frac{x_1^s}{1+\e^{-x_1+y}} \cdot
\frac{\left[N_r+(s+1)Fir^\prime(x_1,y,s)\right]^{\frac{-s}{s+1}}
-x_1^{-s}}{N_r-(s+1)Fir(x_1,y,s)} . \label{varepsxp}
\end{eqnarray}

\subsubsection{Low energy expressions for $s\ne 0$}

All the formulas in Section \ref{diless} are very difficult to 
calculate numerically. Therefore, low energy asymptotic expressions 
can be very useful in calculations. 
To find the low energy behavior of the functions involved, 
we calculate the Taylor expansions of the dimensionless 
quantities around $\epsilon=0$ ($x=0$). 
We start with the integrals over Fermi distributions, 
\begin{equation}
Fir(x,y,s) = \int_0^{x}\frac{x_1^{s}\,\rmd x_1}{\e^{x_1-y}+1}
\approx \frac{1}{\e^{-y}+1}\cdot\frac{x^{s+1}}{s+1} 
-\frac{1}{\left(\e^{-y}+1\right)\cdot 
\left(\e^{y}+1\right)}\frac{x^{s+2}}{s+2} \label{firapp} 
\end{equation}
and
\begin{eqnarray}
Fir^\prime(x,y,s) &=& \int_0^{x}\frac{x_1^{s}\,\rmd x_1}{
\e^{y-x_1}+1} \approx 
\frac{1}{\e^{y}+1}\frac{x^{s+1}}{s+1}+\frac{1}{\left(\e^{-y}+1\right) 
\left(\e^{y}+1\right)}\frac{x^{s+2}}{s+2} \nonumber \\
&=& \frac{x^{s+1}}{s+1}- Fir(x,y,s) \, , \label{firpapp}
\end{eqnarray}
for $x\ll 1$. 
Using Eqs. (\ref{firapp}) and (\ref{firpapp}) we get asymptotic 
expressions for all the other variables. The Bose DOS is 
\begin{eqnarray}
&&\sigma_{{\rm B},r}(x,y,s)\approx 
\frac{\left[N_r-\frac{x^{s+1}}{\e^{-y}+1}\right]}{
N_r^{1/(s+1)}\left[1+\frac{1}{s+1}\cdot\frac{x^{s+1}}{N_r(\e^{y}+1)}-
\frac{x}{N_r^{1/(s+1)}}\right]} \label{sigapp1} \\ 
&& \approx N_r^{s/(s+1)}+ \theta(s)x\cdot N_r^{(s-1)/(s+1)} 
- \theta(-s)\frac{x^{s+1}N_r^{-1/(s+1)}}{\e^{-y}+1} \cdot
\left[1+\frac{\e^{-y}}{s+1}\right] \, , \nonumber 
\end{eqnarray}
where $\theta(x)$ is the Heaviside step function. 
At $x=\xb=0$, $\sigma_{{\rm B},r}=N_r^{s/(s+1)}$, so it attains a 
finite value, even if $\sigma$ is either zero or infinite. 
$N_r^{s/(s+1)}$ is the average density of the Fermi gas, from zero to 
$\ef$. If $s>0$, $\sigma_{{\rm B},r}$ is increasing with $x$, while 
for $s<0$, $\sigma_{{\rm B},r}$ is decreasing. This is a 
general result, since already from equation (\ref{boseDOS}) one can 
observe that $\sib$ is monotonic for monotonic $\sigma$. 

Bose energies are:
\begin{eqnarray}
\eb&\approx&(s+1)\kb T \int_0^x \frac{\rmd x_1\,x_1^s}{1+
\e^{-x_1+y}} \cdot 
\frac{\left[N_r+\frac{x_1^{s+1}}{\e^{y}+1}\right]^{1/(s+1)}
-x_1}{\left[N_r-\frac{x_1^{s+1}}{\e^{-y}+1}\right]} \nonumber \\
&=& \frac{\kb T N_r^{-s/(s+1)}}{\e^{y}+1}\left\{
x^{s+1} -\theta(s)\cdot x^{s+2}\frac{s+1}{s+2}
\left(\frac{1}{N_r^{1/(s+1)}}-\frac{1}{\e^{-y}+1}\right)\right. 
\nonumber \\
&+&\left.\frac{\theta(-s)}{2}\frac{x^{2s+2}}{N_r(\e^{-y}+1)}\cdot
\left(1 + \frac{\e^{-y}}{s+1}\right)\right\} \, . \label{eBapp}
\end{eqnarray}
In the lowest order, for both, $s<0$ and $s>0$, $\xb$ have the same 
expression: 
\begin{eqnarray}
\xb&\equiv&\frac{\eb}{\kb T}\approx\frac{ N_r^{-s/(s+1)}}{\e^{y}+1}
\cdot x^{s+1} 
\label{eBappmin}
\end{eqnarray}
By eliminating $x$ from Eqs. (\ref{eBappmin}) and (\ref{sigapp1})
we obtain the asymptotic expression of $\sib(\eb)$, 
\begin{eqnarray}
\sigma_{{\rm B},r}(x,y,s)&\approx& N_r^{s/(s+1)}+ 
\theta(s)\cdot N_r^{\frac{s^2+s-1}{(s+1)^2}}(\e^y+1)^{1/(s+1)}\xb^{1/(s+1)}
\nonumber \\
&&-\theta(-s)\cdot N_r^{\frac{s-1}{s+1}}\e^{y}\cdot
\left[1+\frac{\e^{-y}}{s+1}\right]\xb \, . \label{sigapp2} 
\end{eqnarray}
So $\sigma_{{\rm B},r}$ decreases linearly with $\xb$ for $s<0$, while 
for $s>0$ it has an infinite positive slope at $\xb=0$. 

The expressions for the interaction potential are more complicated. 
Let us start with $x^\prime\ge x$. 
After some calculations we arrive to the following results: 
\begin{eqnarray*}
v(x,x^\prime\ge x) &=& - \frac{s+1}{C(\kb T)^s}\cdot
\frac{N_r^{\frac{-2s-1}{s+1}}}{\e^y+1}
\left[\frac{s x^{s+1}}{s+1}+\frac{x^{s+2}}{s+2}\cdot\left(
\frac{s}{1+\e^{-y}}-\frac{s+1}{N_r^{\frac{1}{s+1}}} \right)\right] ,\\
&& {\rm for}\  s>0 
\end{eqnarray*}
and 
\begin{eqnarray*}
v(x,x^\prime\ge x) &=&  - \frac{1}{C(\kb T)^s}\cdot
\frac{N_r^{\frac{-2s-1}{s+1}}}{\e^y+1}\left\{ 
sx^{s+1}+\frac{2s+1}{2(s+1)}\left(1+\frac{\e^{-y}}{s+1}\right)\right.\\
&&\left.\times\frac{x^{2s+2}}{N_r(\e^{-y}+1)}\right\} ,
\ {\rm for}\ s<0 \, . \nonumber
\end{eqnarray*}
Obviously, $v(x,x^\prime\ge x)\to 0$, as $x\to 0$, since $\eb=0$ is the 
ground state for any configuration. 

For $x>x^\prime$, 
\begin{eqnarray}
v(x,x^\prime<x) &=&  -\frac{s+1}{C(\kb T)^s N_r} \cdot
\frac{N_r^{\frac{-2s-1}{s+1}}}{\e^y+1}
\left[(x-x^\prime) + \frac{x^2-(x^\prime)^2}{2(1+\e^{-y})}-
\frac{x^{s+1}}{s+1} \right.\nonumber \\ 
&&\left.+ (x^\prime)^{s+1}\right] ,\ {\rm for}\ s>0 , \label{vxp>xs>0}
\end{eqnarray}
and
\begin{eqnarray}
&& v(x,x^\prime<x) = 
\frac{N_r^{\frac{-2s-1}{s+1}}}{C(\kb T)^s(\e^y+1)}
\left[x^{s+1}-(x^\prime)^{s+1}-(x-x^\prime)\cdot(s+1)\right. \nonumber \\
&& \left.\times\frac{x^{2s+2}-(x^\prime)^{2s+2}}{2(s+1)N_r(1+\e^{-y})}
\left(s+1-s\e^{-y}\right) -s(x^\prime)^{s+1} - 
\frac{2s+1}{2(s+1)}\frac{(x^\prime)^{2s+2}}{N_r(\e^{-y}+1)}
\right] \, , \nonumber \\
&& \ {\rm for}\ s<0 . \label{vxp>xs<0}
\end{eqnarray}
%
%
In both equations (\ref{vxp>xs>0}) and (\ref{vxp>xs<0}) only the two 
lowest orders are correct, whichever these are, depending on $s$. 
Again, note that $v(x,x^\prime<x)\to 0$ as $x\to 0$, for any $s$. 

\begin{figure}[t]
\begin{center}
\unitlength1mm\begin{picture}(80,70)(0,0)
\put(0,0){\epsfig{file=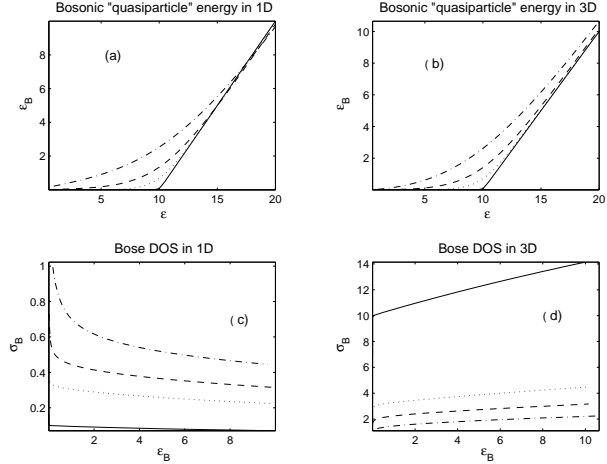,width=80mm}}
\end{picture}
\caption{The Bose quasiparticle energy, $\eb$, vs. $\epsilon$, for (a) 
1D ($s=-1/2$) and (b) 3D ($s=1/2$) systems, and the Bose DOS, $\sib(\eb)$ 
for the same systems. In all the cases we take $\ef=10\kb T_0$, where 
$T_0$ is a scaling temperature. The four curves in each plot correspond to 
four values of $T$: $0.1T_0$ (solid line), $T_0$ (dotted line), $2T_0$ 
(dashed line), and $4T_0$ (dash-dot line).}
\label{fig_eBsiB}
\end{center}
\end{figure}

In Figs. \ref{fig_eBsiB} (a) and (b) it is shown the dependence of 
$\eb$ on $\epsilon$ for 1D and 3D gases and four different temperatures. 
The dependence of $\sib$ on $\eb$ is shown in figure \ref{fig_eBsiB} 
(c) and (d) for the same gases and temperatures. 
The four curves in each of the plots correspond to four different 
temperatures, namely $0.1T_0$, $T_0$, $2T_0$, and $4T_0$, where 
$T_0$ is a scaling temperature and $\ef$ is chosen so that 
$\ef=10\kb T_0$. 

\subsection{THE EQUIVALENCE OF THE BOSE AND FERMI DESCRIPTIONS \label{inequivalence}}

\begin{figure}[t]
\begin{center}
\unitlength1mm\begin{picture}(123,50)(0,0)
\put(5,3){\includegraphics[width=55mm]{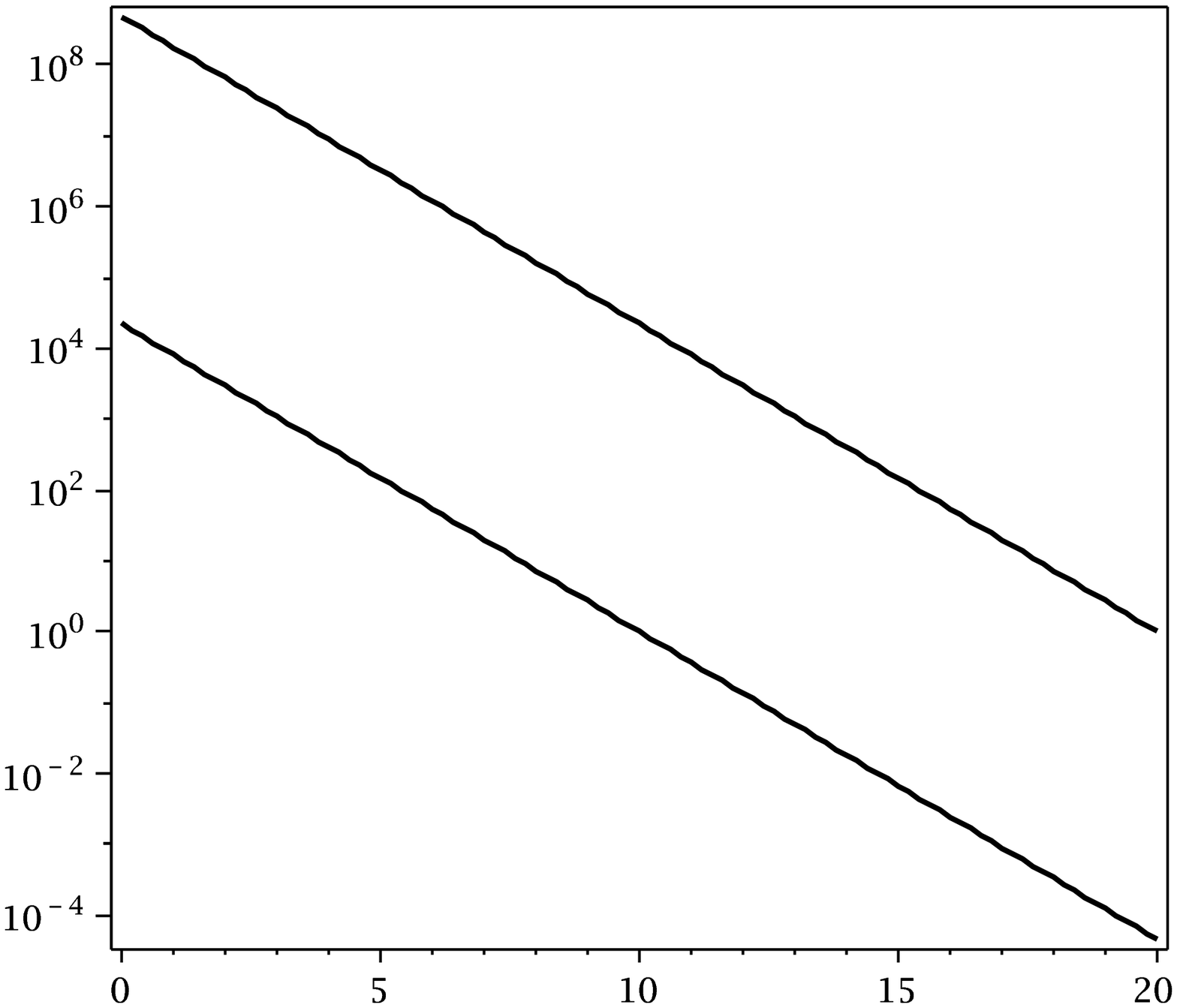}}
\put(70,3){\includegraphics[width=55mm]{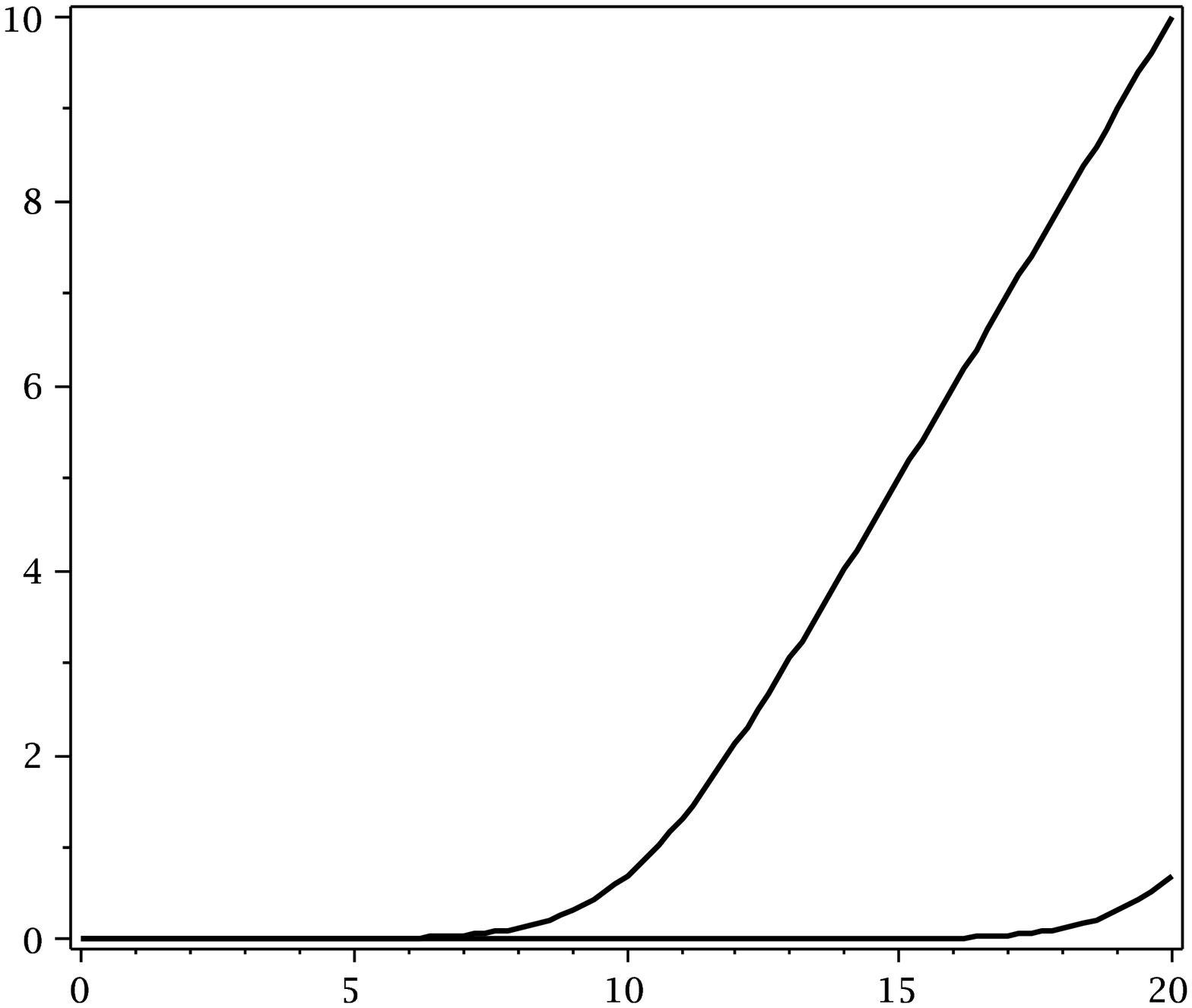}}
\put(50,43){(a)}\put(80,43){(b)}
\put(22,22){1}\put(27,33){2}\put(110,30){1}\put(120,10){2}
\put(0,30){$b_{\rm B},$}\put(0,25){$b_{\rm F}$}\put(34,0){$\epsilon/(\kb T)$}
\put(65,25){$\frac{\bar\epsilon_{\rm B}}{\kb T}$}\put(98,0){$\epsilon/(\kb T)$}
\end{picture}
\caption{For a 1D Fermi gas ($s=-1/2$): (a) The population of the bosonic 
energy levels, 
$b_{\rm F}[\bar\eb(\epsilon)]$ and $b_{\rm B}[\bar\eb(\epsilon)]$, 
for $y=\mu/(\kb T)=10$ (1) and $y=\mu/(\kb T)=20$ (2). For both values of $y$, 
the curves corresponding to $b_{\rm F}$ and $b_{\rm B}$ are indistinguishable. 
(b) The bosonic effective quasiparticle energy, $\bar\eb(\epsilon)$, for the 
same values of $y$ as in (a).}
\label{muB0}
\end{center}
\end{figure}

Having now the bosonic picture of our Fermi system, built based on the 
Fermi distribution, we can analyze the bosonic ensemble as in 
Section \ref{eq_distr}. Using the results 
of Section \ref{ncDOS} we calculate $b$ by both formulas (\ref{bosepopt}) 
and (\ref{bosedistr}) and compare the results. 
To avoid confusion, we denote by $b_{\rm F}$ and $b_{\rm B}$ the 
populations given by equations (\ref{bosepopt}) and (\ref{bosedistr}), 
respectively. 
Since the two systems are thermodynamically equivalent, $b_{\rm F}$ and 
$b_{\rm B}$ should be identical in the thermodynamic limit. This 
identity appears to be difficult to prove analytically, so we 
calculated numerically $\bar\eb(\epsilon)$ and with this, $b_{\rm B}$, 
which we then compared with $b_{\rm F}$. As one can see in figure \ref{muB0}, 
$b_{\rm F}[\bar\eb(\epsilon)]$ and $b_{\rm B}[\bar\eb(\epsilon)]$ are 
equal within the numerical accuracy, for any $\mu/(\kb T)$. (Note: we 
plotted the numerical results for $s=-1/2$ because for this value of 
$s$ the results are much more accurate and the multiple integrals 
involved converge for a wider range of chemical potentials and single particle 
energy levels.)

%

\section{CONCLUSIONS \label{concl}}

In this paper it is introduced the basic idea and results of what have 
been called {\em exclusion statistics transformation} (EST). 
The EST from Fermi to Bose systems is done in the following way: 
the fermions are pictured as piles of particles 
that occupy completely intervals along the single particle energy axis 
(no empty single particle states in each of the intervals), which are 
then rotated into horizontal position by associating to all of them 
a single ``Bose'' energy (see figure \ref{total_B}).
Using EST we transformed an ideal Fermi gas into a Bose gas with the same 
excitation spectrum. Having the same excitations, the two gases are 
thermodynamically equivalent by construction \cite{PhysRevE.55.1518.1997.Lee}, 
i.e. if the Fermi gas has entropy $S_{\rm F}$, internal energy 
$U_{\rm F}$ and particle number $N$, then the Bose gas has the same 
entroy at any temperature as the Fermi gas, $S_{\rm B}=S_{\rm F}$. 
The internal energy of the Bose gas is $U_{\rm B}=U_{\rm F}-U_{\rm g.s}$, 
where $U_{\rm g.s.}$ is the energy of the Fermi system at zero 
temperature. 

Having constructed in this way the Bose gas, one can calculate all the 
thermodynamic quantities of the Bose and Fermi gases independently, 
by maximizing the partition functions with respect to the quasiparticle 
levels populations. Based on the thermodynamic equivalence, all the 
canonical properties of the two gases should be identical and, moreover, 
applying EST, the Fermi distribution should transform into the Bose 
distribution and vice-versa. We checked this identity numerically, by 
calculating the Bose and Fermi populations. The populations 
transform indeed into each-other by EST within the numerical accuracy. 
So the thermodynamic equivalence is not applicable only to a very special 
class of systems, namely ideal systems with the same, constant, DOS, 
but should be regarded as a very general concept. We have now a 
method to transform gases of a given exclusion statistics into 
equivalent gases of some other exclusion statistics. 

Although this is not discussed here, at low temperatures the 
Bose system condenses and the Bose condensate 
is related by EST to the particles in the Fermi system that stay on the lowest 
energy levels and occupy {\em completely} an energy interval. By analogy 
to the Bose system, these particles form the so called Fermi condensate 
\cite{JPA36.L577.2003.Anghel}. 
Using this analogy, the Fermi system may be described as consisting of 
a condensate of, say $N_0$ particles, and with $N-N_0$ particles above the 
condensate. The condensate forms in interacting systems Ref. 
\cite{JPA35.7255.2002.Anghel,RomRepPhys59.235.2007.Anghel}), but also 
in ideal systems 
\cite{JPA36.L577.2003.Anghel}. At low temperatures, 
when the condensate lies close the the Fermi energy, the single particle 
spectrum of the ``thermally active'' $N-N_0$ particles may be approximated 
as constant and the EST becomes very simple (in any number of dimensions), 
since it reduces to the EST in systems of constant DOS 
\cite{JPA35.7255.2002.Anghel}. 

It is well known from Bose systems that the properties of the condensate are 
strongly influenced by the dimensionality of the system (see for example 
\cite{PhysRevLett.84.2551.Petrov}), but the collective properties of 
the condensate, which certainly deserve investigation, were omitted in 
this paper. 

\section{REFERENCES}


\begin{thebibliography}{10}

\bibitem{ProcCambrPhilos42.272.1946.Auluc}
F.~C. Auluck and D.~S. Kothari.
\newblock {\em Proc. Cambridge Philos. Soc.}, 42:272, 1946.

\bibitem{PhysRev.135.A1515.1964.May}
Robert~M. May.
\newblock {\em Phys. Rev.}, 135:A1515, 1964.

\bibitem{AmJPhys63.369.Viefers}
S.~Viefers, F.~Ravndal, and T.~Haugset.
\newblock {\em Am. J. Phys.}, 63:369, 1995.

\bibitem{PhysLettA212.299.1996.Isakov}
S.~B. Isakov, D.~P. Arovasc, J.~Myrheimd, and A.~P. Polychronakos.
\newblock {\em Phys. Lett. A}, 212:299, 1996.

\bibitem{PhysRevLett.74.3912.1995.Sen}
D.~Sen and R.~K. Bhaduri.
\newblock {\em Phys. Rev. Lett.}, 74:3912, 1995.

\bibitem{PhysRevE.55.1518.1997.Lee}
M.~Howard Lee.
\newblock {\em Phys. Rev. E}, 55:1518, 1997.

\bibitem{PhysRevE.56.4854.1997.Apostol}
M.~Apostol.
\newblock {\em Phys. Rev. E}, 56:4854, 1997.

\bibitem{JPA35.7255.2002.Anghel}
D.~V. Anghel.
\newblock {\em J. Phys. A: Math. Gen.}, 35:7255, 2002.

\bibitem{JPA36.L577.2003.Anghel}
D.~V. Anghel.
\newblock {\em J. Phys. A: Math. and Gen.}, 36:L577--L583, 2003.

\bibitem{RomRepPhys59.235.2007.Anghel}
D.~V. Anghel.
\newblock {\em Rom. Rep. Phys.}, 59:235, 2007.
\newblock cond-mat/0703729.

\bibitem{JPA38.9405.2005.Anghel}
D.~V. Anghel, O.~Fefelov, and Y.~M. Galperin.
\newblock {\em J. Phys. A: Math. Gen.}, 38:9405, 2005.

\bibitem{PhysRevB.60.6517.1999.Murthy}
M.~V.~N. Murthy and R.~Shankar.
\newblock {\em Phys. Rev. B}, 60:6517, 1999.

\bibitem{AnnPhysNY270.198.1998.Holthaus}
M.~Holthaus, E.~Kalinowski, and K.~Kirsten.
\newblock {\em Ann. Phys. (NY)}, 270:198, 1998.

\bibitem{AnnPhysNY276.321.1999.Holthaus}
M.~Holthaus and E.~Kalinowski.
\newblock {\em Ann. Phys. (NY)}, 276:321, 1999.

\bibitem{PhysRevLett.84.2551.Petrov}
D.~S. Petrov, M.~Holzmann, and G.~V. Shlyapnikov.
\newblock {\em Phys. Rev. Lett.}, 84:2551, 2000.

\end{thebibliography}

\end{document}